\UseRawInputEncoding 
\documentclass[conference]{IEEEtran}
\usepackage{amsmath}
\usepackage{amsmath,amssymb,amsfonts}
\usepackage{cite}
\usepackage{mathtools}
\usepackage[binary-units=true]{siunitx}
\DeclareSIUnit{\belmilliwatt}{Bm}
\usepackage{amsfonts}
\usepackage{amssymb}
\usepackage{graphicx}
\usepackage{dsfont}
\usepackage{subcaption}
\usepackage{comment}
\usepackage{algorithm}
\usepackage{algpseudocode}
\usepackage{mathtools}

\usepackage{textcomp}
\usepackage[export]{adjustbox}
\usepackage{float}
\usepackage{booktabs}
\usepackage{multicol}
\usepackage{xparse}
\usepackage{color,soul}
\usepackage[bottom]{footmisc}
\usepackage[binary-units=true]{siunitx}
\renewcommand{\hl}[1]{#1}
\DeclareSIUnit{\belmilliwatt}{Bm}
\DeclareSIUnit{\dBm}{\deci\belmilliwatt}
\DeclareSIUnit[per-mode=symbol,per-symbol=p]{\Bps}{\byte\per\second}

\sethlcolor{yellow}
\usepackage{algpseudocode}
\DeclareMathOperator{\E}{\mathbb{E}}
\usepackage{hyperref}
\def\BState{\State\hskip-\ALG@thistlm}
\def\BibTeX{{\rm B\kern-.05em{\sc i\kern-.025em b}\kern-.08em
		T\kern-.1667em\lower.7ex\hbox{E}\kern-.125emX}}

\begin{document}
\title{A TETRA-based System for Remote-Health Monitoring of First Responders: Peak AoI Assessment in Direct and Trunked Mode}

\author{
\IEEEauthorblockN{Hossam Farag\IEEEauthorrefmark{1}, Milo\v s Kosti\' c\IEEEauthorrefmark{3}, Aleksandar Vuji\' c\IEEEauthorrefmark{2}, Goran Bijeli\' c\IEEEauthorrefmark{4}, and \v{C}edomir Stefanovi\'{c}\IEEEauthorrefmark{1}}
\IEEEauthorblockA{
\IEEEauthorrefmark{1}Department of Electronic Systems, Aalborg University, Denmark\\
\IEEEauthorrefmark{3}Tecnalia Serbia, Serbia\\
\IEEEauthorrefmark{2}TeleGroup, Serbia \\
\IEEEauthorrefmark{4}Tecnalia Research and Innovation, Spain\\
Email: hmf@es.aau.dk, milos.kostic@tecnalia.com, aleksandar.vujic@telegroup-ltd.com,\\ goran.bijelic@tecnalia.com, cs@es.aau.dk}}
\maketitle

\begin{abstract}
In this paper, we study peak age of information (PAoI) performance of a novel IoT solution for remote health-monitoring of first responders over TErrestrial Trunked RAdio (TETRA) links.
The solution features a set of sensors embedded in a smart garment that periodically record and send physiological parameters of first responders to a remote agent.
The received data is analyzed by the remote agent, which feeds back notifications and warnings to the first responders in the form of electrotactile stimuli.
The communication in the system is performed over the TETRA Short Data Service (SDS), which is the default option for the development of third-party applications and which has rather limited capabilities. 
The choice of the PAoI as the parameter of interest is motivated by its suitable to measure data freshness in IoT applications with periodic monitoring. 
We derive closed-form expressions of PAoI for different \hl{packet-management} schemes allowed by the TETRA standard, 
and verify the analytical results through extensive simulations under varying message generation rates.
Our results provide important insights on the expected PAoI performance, which can be used for the system design guidelines.
To the best of our knowledge, this is the first work that analyzes AoI performance of TETRA networks.
\end{abstract}

\begin{IEEEkeywords}
AoI, remote health-monitoring, first responders, TETRA
\end{IEEEkeywords}

\maketitle
\section{Introduction}
\label{sec:introduction}
\IEEEPARstart{W}{ith} the proliferation of smart connected devices, the Internet of Things (IoT) technology is gaining a significant attention from both academic and industrial communities~\cite{IoT-1}.
A typical IoT system combines sensors, wireless networking, and remote analytic, 
which is a versatile platform for implementation of a wide range of applications.
Remote-health monitoring involving automated data analysis and decision making is a showcase example of such IoT system.
Employing such system at a rescue site can help first responders to 
do their duty and perform an ongoing operation in a safe and efficient way.
Therefore, public safety agencies have recently commenced research and development projects aiming to employ innovative IoT technologies to improve the safety, situational awareness and operational effectiveness for first responders~\cite{NGFR}.

The public safety networks standardly employ professional mobile radio-systems, where by far the most commonly used is the TErrestrial Trunked RAdio (TETRA)~\cite{TETRA}.
\hl{TETRA systems have been primarily designed and used to support reliable and secure voice communication for public safety and commercial sectors. However, the use of TETRA network in emerging data-oriented and mission-critical applications (which is the case of our considered remote health monitoring scenario) has not been studied before. This is to the best of our knowledge, obtained after a comprehensive literature study.}
Remote health monitoring is characterized by sporadic, quasi-periodic communication of short packets delivering measurements of the monitored physiological parameters in the uplink and providing feedback about the physiological strain in the downlink.
In effect, the required data rates are expected to be rather small.
In the context of the TETRA standard, such low-rate data communications can be supported through Short Data Service (SDS)~\cite{TETRA-AIR}, which is effectively the only option for data communications that is practically available in real-life TETRA deployments.
SDS is fairly similar to short message service in GSM, offering transmission of messages of up to 2047~bits in size.
The transmission of SDS data takes place over a shared signalling channel, which imposes challenges with respect to providing reliable and real-time communication.
On the other hand, it is critical that the command center maintains real-time monitoring of physiological parameters of the first responders for timely detection of physiological disorders.
This requirement can be expressed in terms of the age of information (AoI)~\cite{AoI-concept}, a recently established metric that is particularly suitable for the scenarios that involve quasi-periodic monitoring and control.
Specifically, AoI is the time elapsed since the moment of the generation of the last successfully received update.
In the context of remote-health monitoring application, AoI will express how `fresh' are the measures of the physiological parameters of a first responder that the command centre is currently handling. \hl{In that sense, AoI, as a process-level metric, is the suitable metric to capture information freshness for such scenario compared to delay which is a packet-level metric. For instance, lower  packet delay could be achieved by reducing the update rate, however, this also lead to the monitoring side having outdated status information. Hence, the goal of achieving timely updating is not the same as minimizing the delay, which necessities the need to investigate the AoI performance of remote health-monitoring scenario. In fact, the TETRA technology poses some limitations and challenges for data-oriented applications due to the low bandwidth that is established over the TETRA random access channel, as well as specific packet management possibilities. Such connectivity configurations, along with a realistic setting of use-case scenario, have not been investigated in the existing theoretical works that address the AoI where more abstract and high-level models are typically considered.}

In this paper, we present an assessment of the Peak AoI (PAoI) performance of TETRA network as the connectivity enabler for remote health monitoring of first responders. We show that an adopted packet management has a significant effect on improving the AoI performance, in particular for high message rates, enabling timely monitoring of the first responders' health status. Our results provide insights on the expected performance of TETRA SDS when adopted to the considered remote-health monitoring scenario, which could be used for optimizing the key system parameters for the target performance requirements.
The paper is an AoI-focused extension of our previous work~\cite{WCNC2022}, featuring the following added contributions:
\begin{itemize}
    \item We investigate the performance of the proposed health monitoring scenario in both TMO (Trunked Mode of Operation) and DMO (Direct Mode of Operation) under different \hl{packet management} policies.
    \item We derive closed form expressions for the average PAoI under different queuing disciplines that are available in TETRA technology.
    \item We developed a discrete-event simulator to validate the analytical results and evaluate the link-level performance in terms of the average PAoI and packet loss ratio under various network settings. 
\end{itemize}
The remainder of this paper is organized as follows.
Section~\ref{sec:background} briefly reviews public-safety networks and related work.
Section~\ref{sec:architecture} introduces a high-level view of the system architecture of the remote-health monitoring application.
Section~\ref{sec:TETRA-SDS} illustrates the procedures of TETRA SDS transmission over TMO and DMO links.
Section~\ref{sec:model} gives analytical analysis of the PAoI for different \hl{packet-management} disciplines.
Section~\ref{sec:evaluation} presents the evaluation methodology and the results, and finally the paper is concluded in  Section~\ref{sec:conclusions}.

\section{Background and Related Work}
\label{sec:background}

TETRA is the European Telecommunications Standard Institute (ETSI) standard that was specifically designed for emergency services, public safety and military usage, to provide robust and secured communication even under disaster conditions.
The set of TETRA services include voice calls (group or private), circuit-switched data, packet-switched data and SDS.
The services are characterized by mission-critical performance such as fast call setup fast message transmission, priority-based call handling, advanced encryption, and authentication.
A TETRA system can handle communications over two modes of operation, Trunk Mode Operation (TMO) and Direct Mode Operation (DMO). In TMO, service takes place over the switching infrastructure composed of the TETRA Base Station (BS)~\cite{TETRA-AIR}.
In DMO, TETRA terminals can directly communicate to each other when they are out the coverage of the TETRA BS~\cite{TETRA-DMO}.
The TETRA DMO terminals then connect to the TETRA BS via  gateway terminal (DM-gateway), which is within the TETRA coverage and acts as a relay for the DMO terminals~\cite{TETRA-gateway}. 

The TETRA network for public safety users operates primarily in the frequency range 380~MHz to 400~MHz with 25~KHz carrier spacing and 10~MHz duplex spacing.
With $\pi/4$ Differential Quadrature Phase Shift Keying (DQPSK) modulation scheme, TETRA can reach long communication ranges at the cost of low data rates where the maximum achievable data rate in the original release of the standard is 28.8~kbit/s~\cite{TETRA-AIR}.
The later version of the standard introduced TETRA enhanced data services~\cite{TEDS} with data rates of up to 500~kbit/s.
The TETRA standard~\cite{TETRA-PEI} defines PEI interface for connecting peripheral data devices to TETRA mobile terminals while integrating AT commands and IP connectivity with full access to resources on the network side.
Although PEI offers a number of service options~\cite{TETRA-PEI,TETRA-SCADA},  manufacturers typically provide access only to the SDS~\cite{TETRA-SDS-perf}.
This service has limited capabilities and provides no performance guarantees; its  main advantage is that it is accessed using simple AT commands.

There are other relevant standards for public safety communications, such as P25 and  Digital Mobile Radio (DMR)~\cite{DMR}, featuring a lot of analogies to TETRA.
Another recent competitor in the public safety niche is LTE.
LTE networks offer greater bandwidth and flexibility when it comes to data-rich applications and video, enabling emergency services, first responders and military organisations to integrate data seamlessly with their voice communications.
LTE introduced relevant features and services (e.g., Push-To-Talk, side-link mode that is analogous to DMO in TETRA) starting from Release 13~\cite{Rel13}.
In that sense, a major advantage of LTE is the native support for broadband data services and the commercial availability of LTE modems and development boards that enable straightforward implementation of customized  applications.
However, the commercially available LTE devices supporting side-link mode, which is the analogue of TETRA's DMO, are primarily foreseen for automotive use cases, featuring large form factors and high energy consumption, which makes them unsuitable for the use-cases involving first responders.
    \begin{figure*}[t!] 
		\centering
		\includegraphics[width=0.75\textwidth]{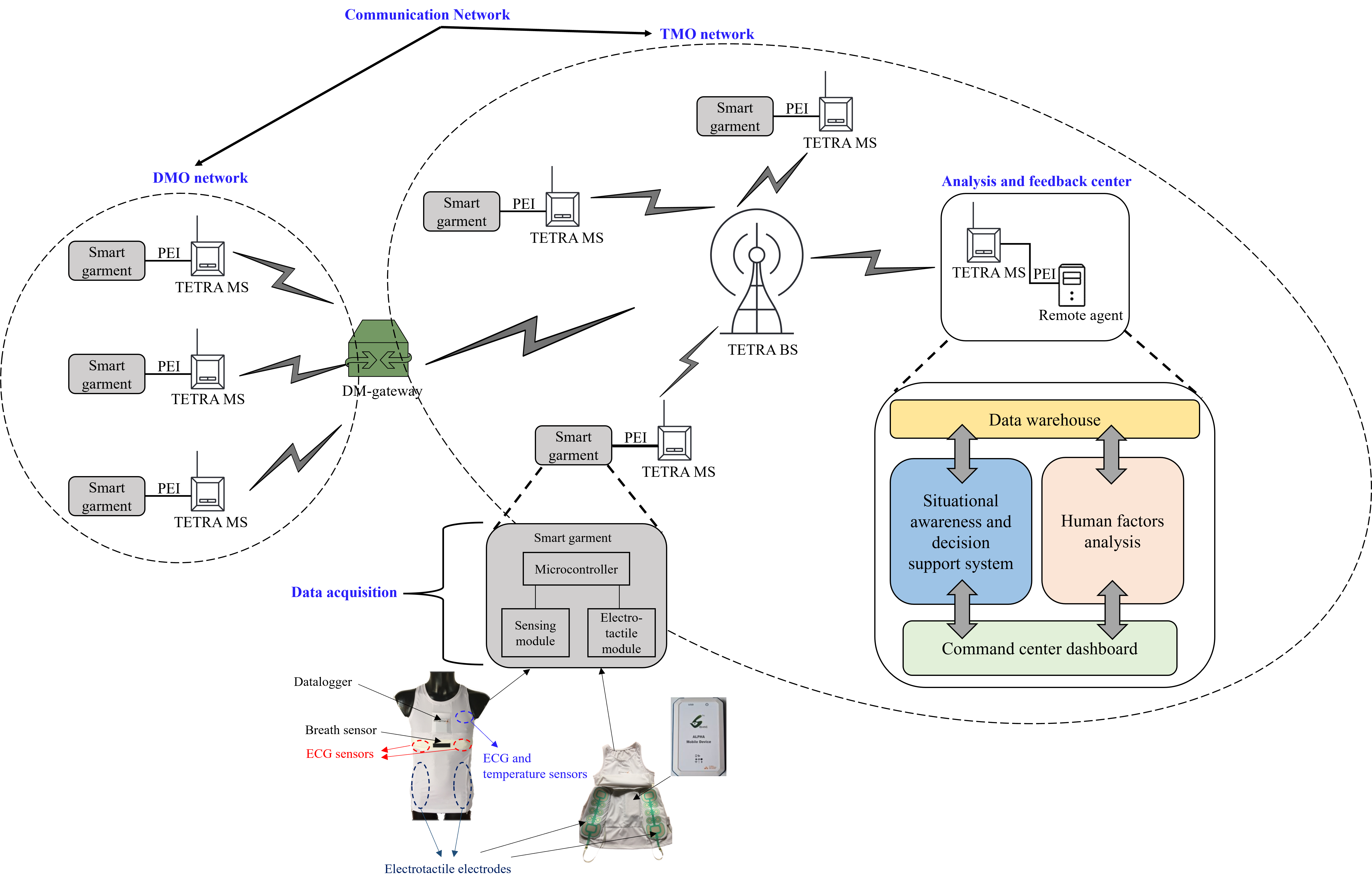}
		\caption{\hl{Block diagram of the considered remote health-monitoring system.}}
		\label{fig:system}
	\end{figure*}

There are a limited set of prior works that evaluate the performance data communication in TETRA under different design parameters ~\cite{TETRA-SCADA,TETRA-Power, TETRA-SDS-perf, TETRA-GPS, TETRA-SDS-perf-DMO}.
The authors in~\cite{TETRA-SCADA} introduce a protocol converter designed for the SCADA system of used in a gas distribution network.
The proposed protocol is based on the  TETRA SDS service to connect remote SCADA units to the supervision center where SDS messages are sent and received with  AT commands.
The work in ~\cite{TETRA-Power} investigates the use of TETRA SDS service for remote monitoring and controlling of substations in the electrical distribution network.
The results show that for a 12~byte SDS message, the system can achieve an average turnaround time of 0.8~s, which would further increase by 12~ms for each additional byte of message size.
The authors in~\cite{TETRA-SDS-perf} evaluate the performance of TETRA SDS-transport layer in terms of delay and message outage among two Mobile Stations (MSs) for different message sizes and transmission intervals.
Their results show that the average delay is between 0.4~s and 1.6~s for message sizes between 10~bytes and 190~bytes.
Moreover, the message outage was negligible for the transmission intervals of 1.5~s and 2~s, while it increases to 1.6\% when the transmission interval reduces to 1~s with more performance degradation when the number of MSs increases to 4.
The study in~\cite{TETRA-GPS} presents performance measurements of the round-trip times of TETRA SDS messages generated by two MSs under static and mobility scenarios. 
In ~\cite{TETRA-SDS-perf-DMO} the authors perform an empirical study of the SDS performance considering TETRA DMO in Vehicular Ad-hoc Network (VANET) supporting safety related railway traffic applications. The measurements campaign revealed SDS transmission delay between 200~ms and 700~ms which show that TETRA can be adopted as a VANET in the railway environment, and those applications requiring reliable direct train-to-train communication. \hl{All the aforementioned works study the performance of TETRA in terms of delay and reliability. However, such metrics are not sufficient to reflect information freshness where the remote agent must have timely status updating of first responders.}

The analysis of the AoI in networked systems has been recently attracting a lot of research interest, in particular for the systems with real-time updates.
The work in~\cite{AoI-FCFS} represents the first queuing theoretic work that studied the average AoI of a single-source First-Come-First-Served (FCFS) queuing model.
The authors in~\cite{AoI-multisource} extended the basic model to a multi-source scenario and derived the exact expression for the average AoI for FCFS M/M/1 queuing model and an approximated expression for the FCFS M/G/1 queuing model.
In~\cite{AoI-Q-management}, the average AoI was derived for different \hl{packet-management} schemes, and a key outcome was that packet replacement can promote reduced AoI compared to the conventional FCFS approach.
The PAoI was introduced in ~\cite{Peak-AoI} as a more tractable metric to evaluate information freshness.
The authors in \cite{Peak-AoI-Prempt} derived closed form expressions for the average AoI and PAoI of a multi-source system under M/G/1/1 preemptive queuing model, however they did not consider the presence of channel errors.
In~\cite{AoI-retransmit}, the authors derived the PAoI of M/M/1 systems considering packet delivery errors and retransmissions. \hl{Other works}\cite{sched1, sched2, sched3}\hl{ have addressed the problem of AoI minimization through different channel access and scheduling policies. However, these are not applicable in our considered system scenario as TETRA follows the standardized random access procedures specified by the ETSI standard} \cite{TETRA-AIR}.  Finally, to the best of our knowledge, there are no prior works addressing AoI in a remote health monitoring system over TETRA links. 

\section{System Architecture and Requirements}
\label{sec:architecture}

The design, development and implementation of the studied system is the topic of the currently running  Horizon 2020 project SIXTHSENSE~\cite{SixthSense}.
Fig.~\ref{fig:system} shows the architecture of the considered remote-health monitoring solution.
It is a closed-loop monitoring system that allows first responders in hazardous situations to sense their current health status through tactile biofeedback induced by transcutaneous stimuli.
Specifically, the first responders wear a smart garment with embedded unobtrusive biosensors that ensures affirm skin contact to accurately record physiological parameters of the first responders, such as temperature, heart rate and heart-rate variability, bio-markers for fatigue, stress, dehydration, etc.
The garment also includes a custom-designed microcontroller that is connected to the TETRA Mobile Station (MS)  via the PEI, where the type of connection between the microcontroller and PEI depends on the manufacturer/device series and can be RS-232, USB, Bluetooth, etc. The microcontroller collects the physiological readings and delivers them to the TETRA MS, which are then sent to the remote agent using SDS service over the TETRA radio links.\footnote{An alternative is to connect to the remote agent, located in some public network (e.g. public Internet) through a Demilitarized Zone (DMZ) of the TETRA network, however, this solution is out of scope of this paper.} When the TETRA MS of the first responder is within the TMO network (i.e., the incident site is located within coverage of the TETRA BS), it can transmit the SDS message directly to the TETRA BS via TMO link, then the TETRA BS forwards the received SDS message to the remote agent. The remote agent receives the relevant SDS message (physiological measurements) in real-time via PEI-based connection to a TETRA~MS. When the TETRA MS is out of coverage of the TMO network, it cannot transmit directly to the TETRA BS, instead, it transmits to the DM-gateway via DMO link. The DM-gateway, which is within the coverage of the TETRA TMO, acts as a rely terminal that connects the TETRA MSs within the DMO network to the TETRA BS as depicted in Fig.~\ref{fig:system}. The DM-gateway  receives the SDS message from the TETRA MS, and forwards it to the TETRA BS via TMO link, then the TETRA BS delivers the received SDS message to the remote agent. The transmission of SDS messages over the TETRA links follows a random channel access scheme(slotted-ALOHA~\cite{TETRA-DMO, TETRA-DMO}). The random access procedures for TMO links and  DMO links are presented in details in Section~\ref{sec:TETRA-SDS}.

The received physiological biomarkers are analysed and visualised through a mission specific decision support system and the remote agent infers the first-responders' physiological condition through warning and feedback signals, which are also sent via SDS service either via TMO link or DMO link depending on the location of the first responder.
This feedback takes the form of a predefined set of tactile stimuli which drives the microcontroller to incite the electrodes in the garment and convey the feedback to the first responders. Such feedback channel does not interfere with the visual and auditory modalities of the first responders, thus minimally compromising their ability to carry out their mission in complex rescue scenarios. Moreover, the generated sensation is not masked by movements, making them suitable for applications involving a demanding physically activity. In communication terms, the system is characterized by quasi-periodic short-data exchanges, both in the uplink (transmission of the physiological measurements to the remote agent) and downlink (feedback transmission to the first responders). An individual sensor reading is a couple bytes long, thus, the size of application-level messages that aggregate multiple readings is of the order of tens to a hundred bytes. Similarly, the set of tactile stimuli that can be unobtrusively used in the system is limited, so a couple of bytes is enough to encode an individual stimulus.

First responders are professionals who are well trained to act under extreme and dangerous situations.
Therefore, there should be a significant build-up of a physiological stress and long periods before their health becomes jeopardized.\footnote{The main concern is that, in the course of the action, the first responders do not themselves recognize that their health status has become compromised. Indeed, overexertion and physiological stress are the leading causes of undesired and even tragic consequences~\cite{FD}.}
In effect, the monitoring interval of their physiological status can be of the order of tens of seconds, or even minutes.
Moreover, the system performs a continuous monitoring, hence the transmission reliability of an individual message, either uplink or downlink transmission, is of secondary importance.
Specifically, the emphasis is on guaranteeing timeliness on the message-flow level for efficient situational awareness, and conventional performance metrics (e.g., throughput and delay) cannot adequately capture the information freshness.
In this context, AoI is introduced as a new metric to quantify the information freshness at the receiver side.
AoI is defined as the time elapsed since the recent received packet was generated, hence AoI jointly captures the latency in transmitting updates and the rate at which they are delivered. Throughput the paper, we will use the terms fragment, packet and update interchangeably to refer to the transmitted information. In Section~\ref{sec:evaluation}, we evaluate the performance of PAoI, which provides information about the worst age, and it is defined as the maximum value of AoI achieved immediately prior the reception of an update.

\section{TETRA Short Data Service}
\label{sec:TETRA-SDS}
\begin{figure}[t!] 
		\centering
		\includegraphics[width= 0.9\linewidth]{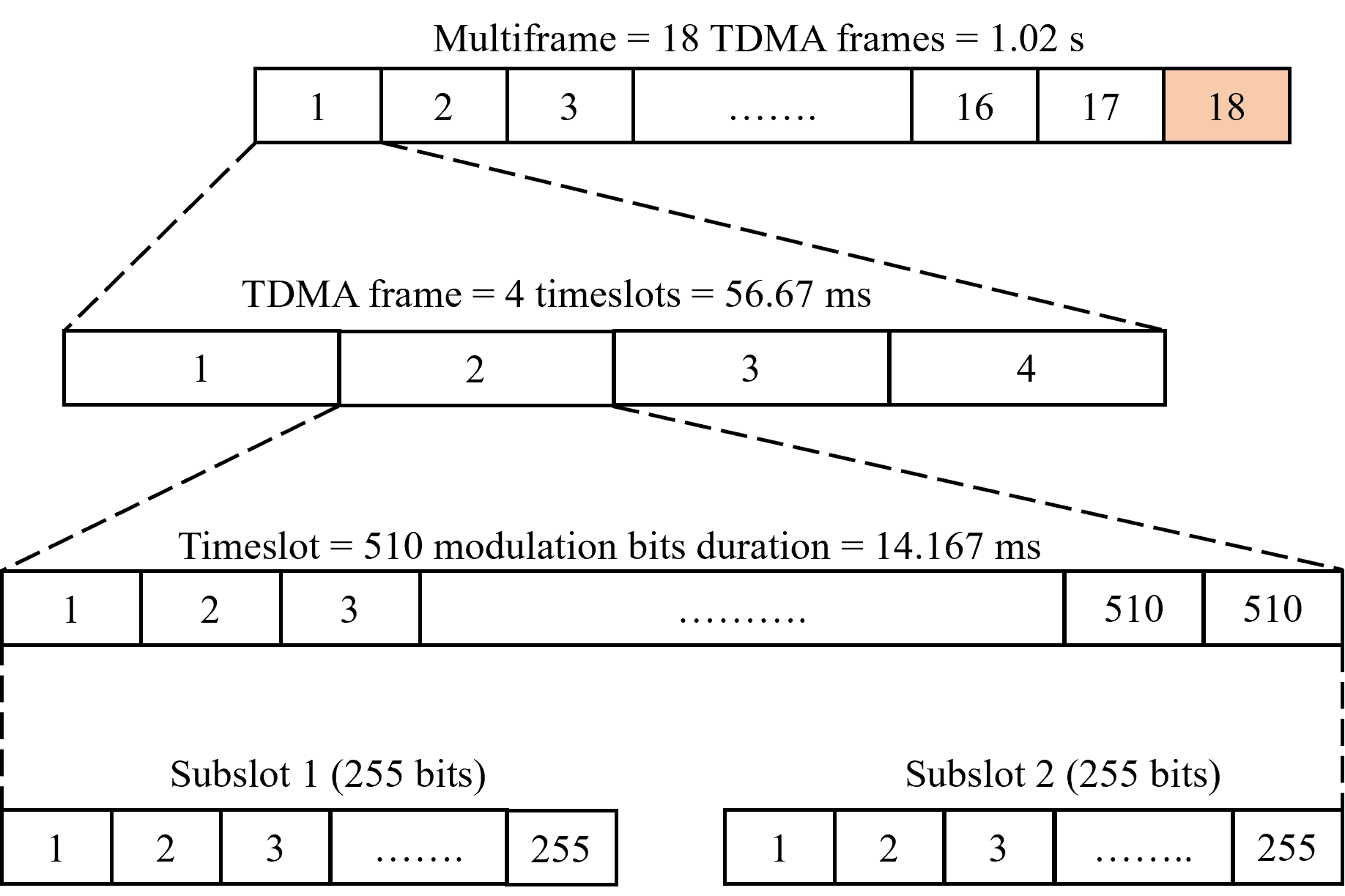}
		\caption{The TDMA structure in TETRA.  \label{tdma}}
	\end{figure}

\hl{The radio sub-system in TETRA is allocated a part of the radio spectrum, which is further partitioned in frequency and time with a carrier spacing of 25~KHz.
The radio access scheme is based on Time Division Multiple Access (TDMA) where time is divided into time slots that are grouped into consecutive TDMA frames as depicted in }Fig.~\ref{tdma}.
\hl{The TDMA structure is composed of multiframes, frames, slots and subslots.
Four time slots form a TDMA frame of duration 57.67~ms.
One multiframe consists of 18~TDMA~frames where the 18th frame is used to carry control signalling.
A physical channel is defined by a pair of radio carrier frequencies (downlink and uplink) and there are two types of  defined by TETRA standard, Traffic Physical channel (TP) which carries voice and data, and Control Physical channel (CP) which carries signaling and SDS messages.
The SDS service provides both pre-defined 16-bit messages and user-defined messages, whose length can be 16 bits (SDS type-1), 32~bits (SDS type-2), 64~bits (SDS type-3) or up to 2047~bits (SDS type-4) of application-defined data} \cite{TETRA-AIR}.
\hl{The sensory information gathered by the microcontroller is placed into the TETRA MS output queue as SDS fragments. Throughput the paper, we will use the terms fragment, packet and update interchangeably to refer to the transmitted information
The SDS messages is sent in fragments if its size exceeds the limit for the MAC PDU.
The TETRA standard defines two modes of operation, Trunked Mode Operation (TMO) and Direct Mode Operation (DMO)~}\cite{TETRA}.
\hl{In the following subsections, we introduce the procedures of SDS transmission in TMO and DMO.} 

\subsection{SDS Transmission in TMO}\label{TMO}

\hl{SDS messages are sent over the shared Main Control CHannel (MCCH) which is located on time slot~1 of the main carrier.} \hl{The MAC layer at each user adopts an invitation-based slotted ALOHA protocol} \cite{TETRA-AIR}. \hl{A downlink broadcast massage, ACCESS-DEFINE message} \cite{TETRA-AIR} \hl{defines the relevant random access parameters, such as the Waiting Time (WT), that denotes the number of TDMA frames before an MS initiates an access retry, ranging from 1 to 15 and the Number of random access transmissions (Nu), that defines the maximum number of random access attempts by an MS, ranging from 1 to 15.}

    \begin{figure}[t!] 
		\centering
		\includegraphics[width= 0.9\linewidth]{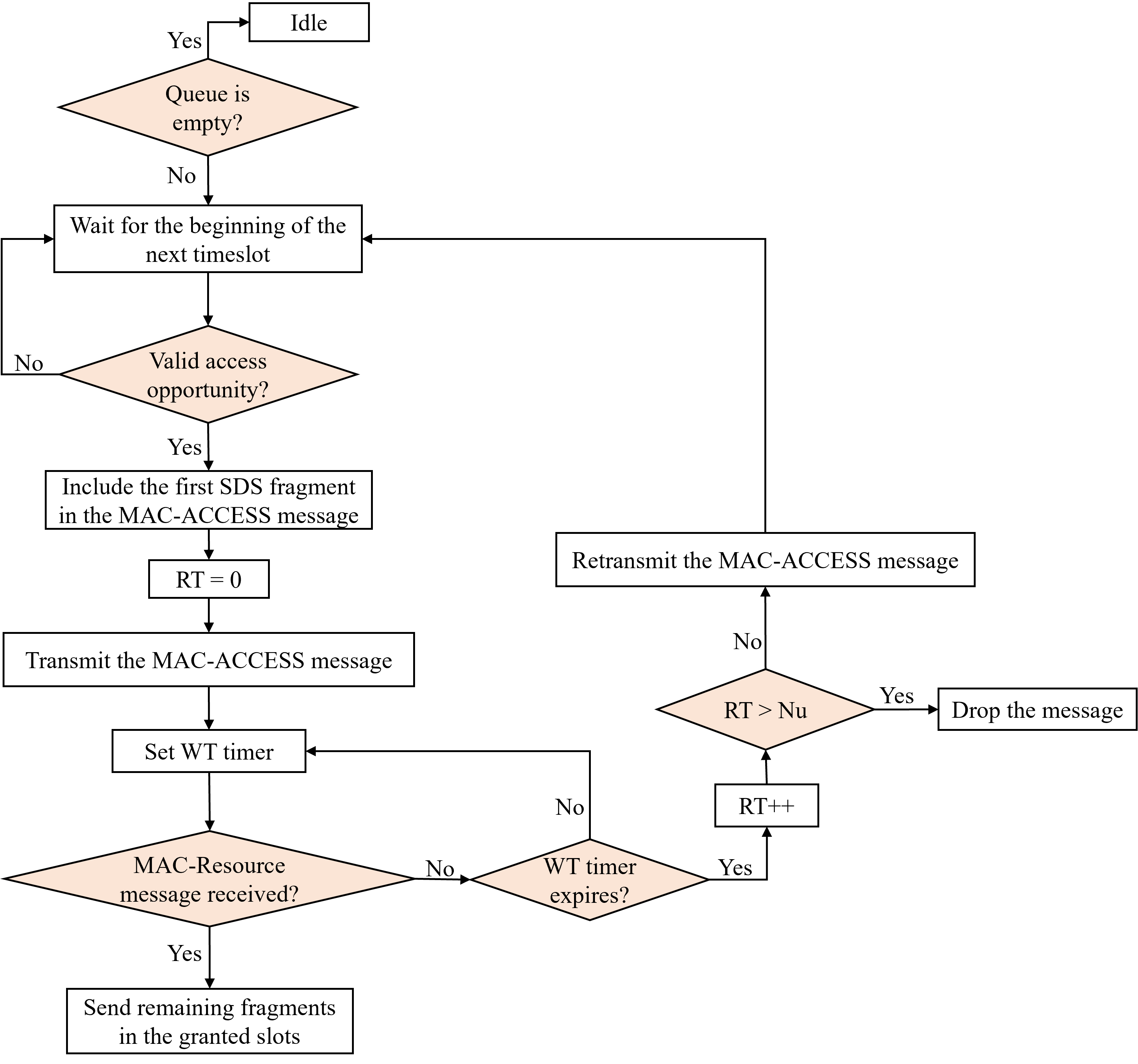}
		\caption{\hl{Flowchart of the random access procedures in TETRA TMO link.}}  \label{RAA}
	\end{figure}
    \begin{figure*}[t!] 
		\centering
		\includegraphics[width=0.8\textwidth]{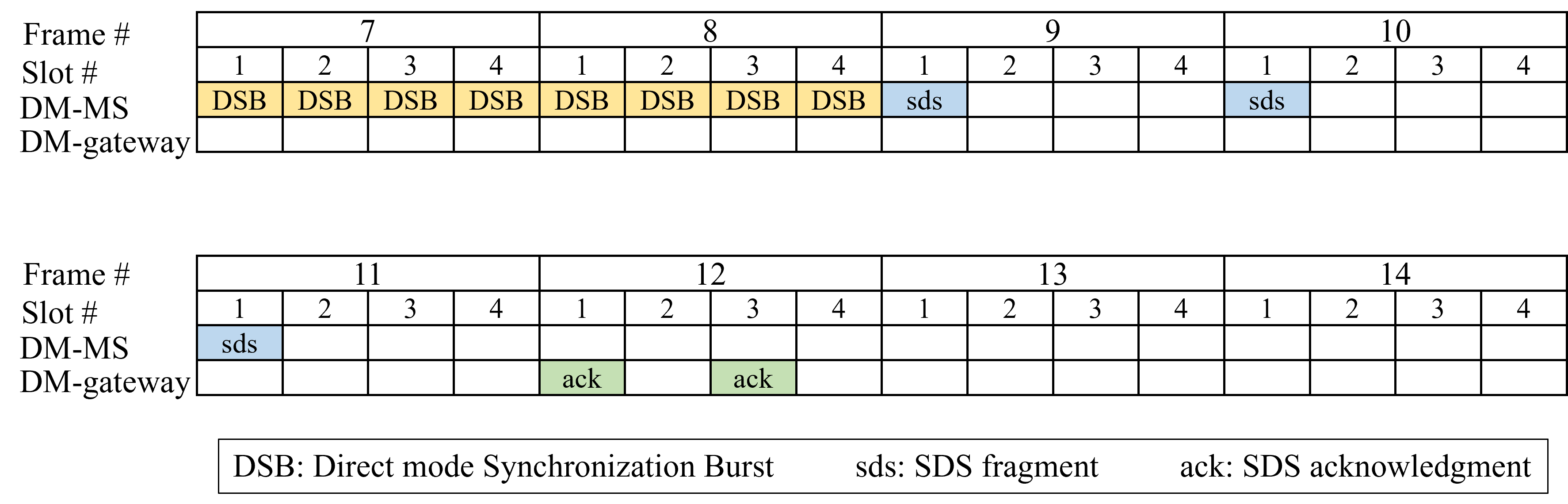}
		\caption{Example of SDS transmission in DMO.  \label{sds-dmo}}
	\end{figure*}
\hl{The random access procedure for transmitting the SDS fragments are described by the flowchart shown in} Fig.~\ref{RAA}\hl{, and explained as follows.
Aligning with a valid access opportunity, the MS transmits a random access request via the  MAC-ACCESS message} \cite{TETRA-AIR} \hl{to the BS.
The MAC-ACCESS message includes the first fragment of the SDS (up to 86~bits} \cite{TETRA-AIR})\hl{ and request for reserved capacity, i.e., number of required transmission opportunities to transmit the SDS fragments.
After sending the MAC-ACCESS message, the user shall wait for the response from the BS which send in downlink via the MAC-RESOURCE message} {\cite{TETRA-AIR}.
\hl{If the response is not received within WT opportunities, the MS shall select randomly another valid access opportunity to retransmit its request.
If the number of retransmission attempts exceeds Nu, the transmission is considered to be failed and the MS falls back to the idle state. 
Otherwise, the user receives the MAC-RESOURCE message from the BS indicating a successful random access and including the corresponding reserved slots. If a single transmission error occurs in one fragment, the whole SDS message fails, and the MS repeats the whole procedure up to a specific retransmission limit.
Upon successfully receiving the fragments and constructing the whole SDS message, the BS sends an acknowledgement (ACK) to the transmitter.
If the ACK is not received within a pre-defined interval, the MS has to retransmit the whole SDS message again.
Upon receiving the SDS message, the BS transmits the received message to the remote agent via reserved access, which is typically granted for downlink transmission in TETRA} \cite{TETRA-AIR}.
\hl{Based on the received sensory information in the SDS messages, the remote agent may in emergency situations transmit the feedback and warnings to the first responders via the BS using a reserved access. 
This is possible because the TETRA BS may reserve a number of transmission opportunity for a particular user when it is required to transmit a solicited message} \cite{TETRA-AIR}. 

\subsection{SDS Transmission in DMO}\label{DMO}

\hl{Direct Mode Operation (DMO) provides direct mobile-to-mobile communications without using the infrastructure in the TMO when the MS is outside the coverage of the TETRA network} \cite{TETRA-DMO}. \hl{An example of SDS transmission in DMO is depicted in} Fig.~\ref{sds-dmo} \hl{and is further illustrated as follows. When the channel is free, the  MS transmits a set of Direct mode Synchronization Bursts (DSB) to establish the channel synchronization and simultaneously its role as master. The DSBs occupy the time slots of a number of consecutive frames (maximum 4 consecutive frames) where the number of signalling frames is defined by frame countdown field in the DSB. Besides the signaling information, the first fragment of the SDS message is included in the DSB. After the synchronization phase, the MS transmits the remaining parts of the SDS message as fragments (sds in } Fig.~\ref{sds-dmo}\hl{) in time slot 1 of each frame. Upon successful reception of the SDS message, the DM-gateway sends acknowledgement (ack in} Fig.~\ref{sds-dmo}) \hl{to the DM MS in time slot 1 and time slot 3 of the next frame. After transmitting the last SDS fragment, the MS starts the timer DT316} \cite{TETRA-DMO} \hl{and waits for the acknowledgment. If the acknowledgment is received within the DT316, the MS stops the timer, otherwise, the MS shall attempt to retransmit the SDS message up to DN316 times} \cite{TETRA-DMO}. \hl{The values of DT316 and DN316 are to be chosen by the MS designer. After the DM-gateway receives the complete SDS message, it will transmit it to the TMO network following the RA procedures described in} Section~\ref{TMO}.

\hl{As mentioned earlier, the timeliness of the exchanged SDS messages is critical for the considered health-monitoring system. However, the standard channel access of TETRA SDS may adversely affects the information timeliness (AoI), especially in dense TETRA networks. Since there is almost no gap to improve the random access in TETRA (it is a standardized procedure), efficient packet management could be a solution to improve the AoI performance of the system. In the next section, we present mathematical analysis of the PAoI considering three different packet-management schemes that can be adopted in TETRA MSs.}    
    \begin{figure*}[t!] 
		\centering
		\includegraphics[width=\linewidth]{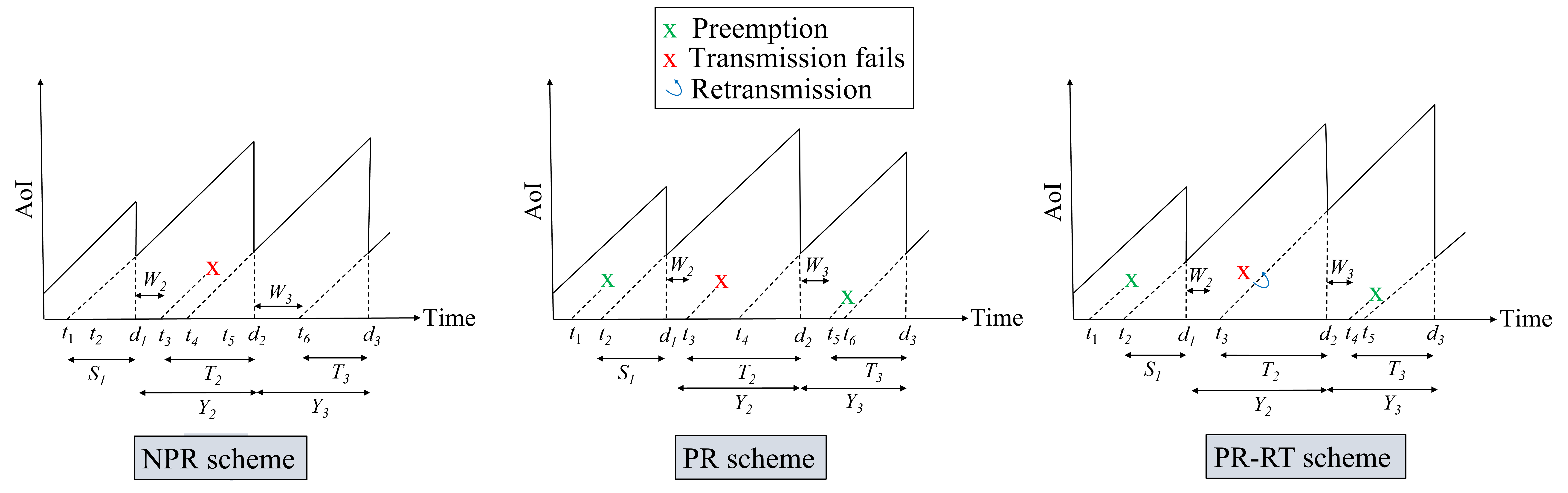}
		\caption{The evaluation of the AoI for NPR, PR and PR-RT.  \label{AoI-model}}
	\end{figure*}
\section{Analysis of the Average PAoI} \label{sec:model}

\hl{PAoI is the maximum value of AoI  immediately before the recent update is received. For mission-critical applications, which is the case of our considered scenario, PAoI is more informative metric than the average AoI where is the interest is to investigate the AoI performance in the worst-case scenario.} In this section, we provide analytical expression for the average PAoI considering three \hl{packet-management} disciplines:
\begin{enumerate}
    \item Preemption without retransmission (PR) scheme: The TETRA standard  allows traffic stealing where an arriving message can preempt the one currently in the service (if there is such) by including a stealing flag in the MAC-ACCESS message header \cite{TETRA-AIR}.
    \item Preemption with retransmission (PR-RT) scheme: Packet preemption is enabled, and the TETRA device of a first responder keeps retransmitting the current update until either the packet is successfully decoded by the remote agent or a new status update is generated, hence preempting the one being retransmitted.
    \item No preemption and no retransmission (NPR) scheme: the device discards any newly generated update while it is busy sending a fragment. In addition, there are no retransmissions.
\end{enumerate}

For the three \hl{packet-management} schemes, we assume that the output buffer of the TETRA MS can accommodate only one packet.\hl{ Our analysis is based on the one presented in} \cite{AoI-retransmit}, \hl{except that we extend the considered model to a single-buffered queue with preemption and retransmission enabled.}

\subsection{Preliminaries}
We assume that the updates (SDS messages) at each first responder are generated according to a Poisson process with intensity $\lambda_F$. We also consider that the generated physiological status can be delivered with a single fragment of the SDS message, which is included within the MAC-ACCESS message (the MAC-ACCESS message can carry a first fragment of 86~bits of information~\cite{TETRA-AIR}).
For PR-RT and PR schemes, we denote $\beta$ as the probability that an update is preempted. The value of $\beta$ is given as a function of the arrival intensity and the service time (see~\eqref{preemp}).  
By $\alpha$ we denote the probability of transmission failure due to the combined effects of channel errors, interference or collisions.
For a successfully received update, the service time $S$ is deterministic for the PR and the NPR schemes and equals to $\mu$ which denotes the duration of a single fragment.
The evaluation of the PAoI at the remote agent for all first responders are statistically identical, hence, in the following we focus on the derivation of the PAoI of an arbitrary user.  

Fig.~\ref{AoI-model} shows an example evolution of the AoI for the three considered queuing schemes, and the corresponding timing parameters are defined as follows.
Let $t_j$, $j=1, 2, 3, ....,$ the generation time of the $j$th update at the first responder.
We denote $X_j$ as the random variable that represents the interarrival time between consecutive updates, $X_j=t_{j+1}-t_j$, which follows an exponentially distribution with mean $1/\lambda_F$. Note that an update may not be received correctly by the remote agent due to transmission failures or preemption. Hereafter, we use a different index $i\leq j$ to refer to the successfully received updates. Let $d_i$ refer to the departure time of $i$th update that is successfully received at the remote agent, and $S_i$ is its corresponding service time. We denote $g_i$ as the generation time of the first generated update after $d_{i-1}$, and is given as
\begin{equation} 
g_i\triangleq \mathrm{min}\{t_j\mid t_j>d_{i-1}\}.
\end{equation}
Therefore, we can see that the indices $i\,\textrm{and}\,j$ in general do not refer to the same update.
For instance, in the PR scheme in Fig.~\ref{AoI-model}, the generated update at $t_3$ is not received by the remote agent, and the successfully received update at $d_2$ is the one generated at $t_4$.
We define $W_i=g_i-d_{i-1}$ as the interval between the reception of the $(i-1)$th update until the generation of the next update. We define the interval $T_i=d_i-g_i$, which represents the interval from $g_i$ until the next update received successfully. Note that $T_i$ spans the generation instants of failed updates. We also define the interdeparture time between two consecutive successfully received status updates $Y_i=d_i-d_{i-1}$. From the definition of $W_i$ and $T_i$, we have $Y_i=T_i+W_i$. Therefore, the PAoI  $A_{i}$ (the value of AoI just before receiving the update at $d_i$), can be given as
\begin{equation} 
A_{i}=Y_i+S_{i-1},
\end{equation}
where $S_{i-1}$ represents the service time of the update received before $d_i$. For instance, in Fig.~\ref{AoI-model}, the PAoI at $d_2$ is equal to $Y_2+S_{1}$, where $S_{1}$ is the service time of the update received before $d_2$, which is $d_1$. The average PAoI, i.e., $\E[A]$, is given by 
\begin{equation} \label{PAoI} 
\E[A]=\E[Y]+\E[S]=\E[T]+\E[W]+\E[S],
\end{equation}
where $\E[Y]=\E[T]+\E[W]$. 

\subsection{Average PAoI for PR Scheme}
For the PR scheme, when a fragment departs, it leaves the system empty, hence $W$ will follow the same distribution as the interarrival time, i.e., $\E[W]=1/\lambda_F$.
Moreover, since there are no retransmissions, we have $\E[S]=\mu$.
The term $\E[T]$ can be evaluated using a recursive method \cite{recursive} as
\begin{equation} \label{T}
\begin{split}
   \E[T]=&\underbrace{(1-\beta)(1-\alpha)\E[S]}_{R_1}\\
   &+\underbrace{(1-\beta)\alpha\left(\E[S]+\E[W]+\E[\hat{T}]\right)}_{R_2}\\
   &+\underbrace{\beta(\E[X \mid X<S]+\E[\hat{T}])}_{R_3}. 
\end{split}
\end{equation}
The first term $R_1$ denotes the case when the first update (generated at $g_i$) is not preempted ($1-\beta$) by other updates and is received successfully by the remote agent. The preemption probability $\beta$ is given as
\begin{equation} \label{preemp}
\beta=1-e^{(-\mu\lambda_F)}.
\end{equation}
The term $R_2$ refers to the case when the first update is not preempted, but the transmission to the remote agent fails.
In this case, the system spends the service time $S$ for the first update, then waits for the period $W$ until the next update is generated. $\E[\hat{T}]$ refers to the interval from the generation of a new update until the receiver correctly receives an update. Note that the evaluation of $\hat{T}$ is the same as $T$, hence $\E[\hat{T}]=\E[T]$. The term $R_3$ represents the case that the first generated update is preempted by a new update. In that case, the effective generation interval, i.e., the generation interval given that a packet is preempted can be given as
\begin{equation} \label{AA}
\E[X\mid X<S]=\frac{\int_{0}^{\mu} s\lambda_F e^{-s\lambda_F }\,ds}{1-e^{(-\mu\lambda_F)}}=\frac{1}{\lambda_F}+\mu\left(1-\frac{1}{\beta}\right).
\end{equation}
Using \eqref{AA} in \eqref{T} and substituting $\E[S]$ and $\E[W]$, $\E[T]$ can be obtained as
\begin{equation} \label{K2}
\E[T]=\frac{\beta+\alpha-\beta \alpha}{\lambda_F(1-\beta)(1-\alpha)}.
\end{equation}
Then, $\E[Y]$ becomes
\begin{equation} \label{Y}
\begin{split}
  \E[Y]&=\E[T]+\E[W]\\
  &=\frac{\beta+\alpha-\beta \alpha}{\lambda_F(1-\beta)(1-\alpha)}+\frac{1}{\lambda_F}=\frac{1}{\lambda_F(1-\beta)(1-\alpha)}.  
\end{split}
\end{equation}
Based on \eqref{PAoI} and \eqref{Y}, we obtain the average PAoI for the PR scheme $\E[A]_{\mathrm{PR}}$ as
\begin{equation} \label{without_ret}
\E[A]_{\mathrm{PR}}=\frac{1}{\lambda_F(1-\beta)(1-\alpha)}+\mu.
\end{equation} 
\subsection{Average PAoI for PR-RT Scheme}
For the retransmission scenario, we need to define $\E[S]$. Let $E_s$ be the average service time when the current update is not preempted by new update and finally received correctly by the remote agent. If we define $p_s$ as the probability of such event, then we have $\E[S]=\frac{E_s}{p_s}$. Then, we define $E_s$ and $p_s$ as follows
\begin{equation} \label{Es}
{E_s}=\sum_{k=0}^{\infty} \alpha^k (1-\alpha)(1-\beta)^{k+1} (k+1)\mu,
\end{equation}
\begin{equation} \label{ps}
{p_s}=\sum_{k=0}^{\infty} \alpha^k (1-\alpha) (1-\beta)^{k+1},
\end{equation}
where $k$ is the number of retransmission rounds. After simplifying \eqref{Es} and \eqref{ps}, we get
\begin{equation} \label{Es2}
{E_s}=\frac{(1-\beta)(1-\alpha)\mu}{(1-\alpha(1-\beta))^2},
\end{equation}
\begin{equation} \label{ps2}
{p_s}=\frac{(1-\beta)(1-\alpha)}{1-\alpha(1-\beta)}.
\end{equation}
Using \eqref{Es2} and \eqref{ps2}, $\E[S]$ can be expressed as
\begin{equation} \label{ESS}
\E[S]=\frac{\mu}{1-\alpha(1-\beta)}.
\end{equation}
For the retransmission case, the evaluation of $\E[T]$ is the same as in \eqref{T} with $\E[S]$ is given in \eqref{ESS}. Then, $\E[Y]$ is expressed as
\begin{equation} \label{YS}
\E[Y]=\frac{1-\alpha+\alpha\beta}{\lambda_F(1-\beta)(1-\alpha)}.
\end{equation}
Using \eqref{ESS} and \eqref{YS}, the average PAoI for the PR-RT scheme $\E[A]_{\mathrm{PR-RT}}$ is given as
\begin{equation} \label{with_re}
\begin{split}
\E[A]_{\mathrm{PR-RT}}=&\E[Y]+\E[S]\\
=&\frac{1-\alpha+\alpha\beta}{\lambda_F(1-\beta)(1-\alpha)}+\frac{\mu}{1-\alpha(1-\beta)}.
\end{split}
\end{equation}
\subsection{Average PAoI for NPR Scheme}
For the NPR scheme, following the same recursive method in \eqref{T}, the term $\E[T]$ is expressed as
\begin{equation}\label{T_NP}
 \E[T]=(1-\alpha)\E[S]+\alpha\E[S+W+\hat{T}]. 
\end{equation}
With $\E[W]=\frac{1}{\lambda_F}$, we have $\E[T]=\frac{\mu+\frac{\alpha}{\lambda_F}}{1-\alpha}$.
Since there are no retransmissions, we have $\E[S]=\mu$.
From \eqref{PAoI}, the average PAoI for the NPR scheme $\E[A]_{\mathrm{NP}}$ is given as
\begin{equation}\label{AoI-NP}
    \begin{split}
  \E[A]_{\mathrm{NPR}}&=\E[T]+\E[W]+\E[S]\\
  &=\frac{\mu+\frac{\alpha}{\lambda_F}}{1-\alpha}+\frac{1}{\lambda_F}+\mu.
    \end{split}
\end{equation}
\begin{table}[t!]
		\centering
		\caption{Evaluation parameters.}
		\label{t1}
		\begin{tabular}{ll}
			\toprule
			Parameter & Value \\
				\midrule
			Modulation & $\pi/4$-DQPSK\\
			Carrier frequency & \SI{400} {\mega\hertz}\\
			$\mathrm{N_C}$ & \num {500}\\
			$\lambda_{c}$ & \num {10} messages/hour/MS\\
			$\lambda_{\textrm{voice}}$ & \num {3} calls/hour/MS\\
			Call duration & Uniformly distributed [\SI{20}-\SI{40}{\second}]\\
			\bottomrule
		\end{tabular}	
	\end{table}
\begin{table}[t]
\centering
 \caption{\hl{packet-management} schemes in different settings.}
 \label{settings}
  \begin{tabular}{|l|c|c|}
  \hline
 & First Responder & DM-gateway \\
  \hline
  Setting (1) & PR-RT & FCFS\\
  \hline
   Setting (2) & FCFS & FCFS\\
  \hline
  Setting (3) & PR-RT & \cite{queu-mm1}\\
 \hline
 \end{tabular}
 \end{table}
    \begin{figure}[t!] 
		\centering
		\includegraphics[width= 1\linewidth]{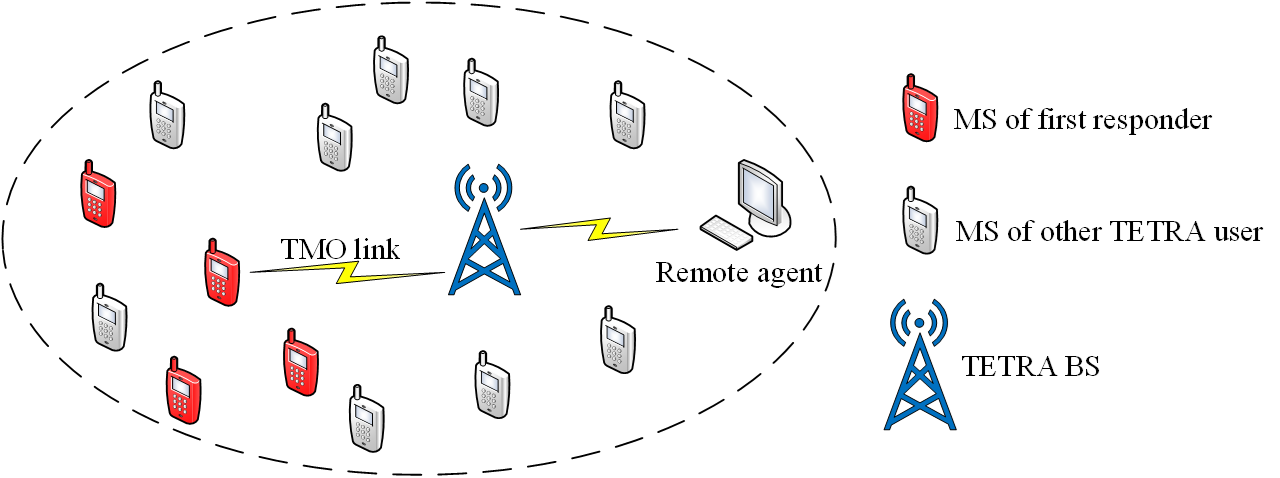}
		\caption{Graphical representation of the simulated TMO model of TETRA network.  \label{netw_TMO}}
	\end{figure}

\section{Evaluation}
\label{sec:evaluation}
In this section, we evaluate the average PAoI performance of end-to-end SDS transmissions in TETRA network under two scenarios.
In the first scenario, the first responders are located within the coverage area of the TETRA BS and connect only using TMO.
In the second scenario, we consider the first responders are located outside the coverage are of the TETRA BS and connect through DMO links via a Direct Mode gateway (DM-gateway)~\cite{TETRA-gateway}.

\subsection{TMO service performance}

First, we validate our analytical model presented in Section~\ref{sec:model} with simulation results obtained from MATLAB implementation.
Then, we introduce simulation assessments that show the impact of different \hl{packet-management} disciplines and network settings on the AoI performance of TETRA.
The results are obtained via a discrete-event simulator in MATLAB that runs the TETRA random access procedures described in Fig.~\ref{RAA}. 

We consider the TETRA TMO network model depicted in Fig.~\ref{netw_TMO}, where there are a total of $\mathrm{N_{tot}} = \mathrm{N_C} + \mathrm{N_F} + 1$ users that are uniformly distributed within the cell coverage area, communicating with the TETRA BS through TMO radio links.
Specifically, $\mathrm{N_C}$ users are standard MSs generating both voice and SDS traffic according to a Poisson process with intensities $\lambda_{\textrm{voice}}$ and $\lambda_\text{C}$, respectively.
We set $\mathrm{N_C}$ to 500, which in practice can be assumed to represent catastrophic situations in which a huge number of public safety forces are concentrated in a small area~\cite{TETRA-users-cell}.
A subset of $\mathrm{N_F}$ users represents the the MSs of first responders, transmitting only SDS messages to the remote agent, following a Poisson process with intensity $\lambda_\text{F}$.
The final user is the remote agent -- a fixed terminal within the cell that generates a 1-byte feedback SDS messages to the $\mathrm{N_F}$ users following a Poisson process with intensity $\lambda_\text{R}=\frac{\mathrm{N_F}}{\SI{60}{\second}}$.
Unless otherwise stated, the parameters listed in Table~\ref{t1} are used in the evaluation; their values are selected according to the ETSI standards~\cite{TETRA, TETRA-AIR}, modeling real-world scenarios. 

    \begin{figure}[t!] 
		\centering
		\includegraphics[width= 0.9\linewidth]{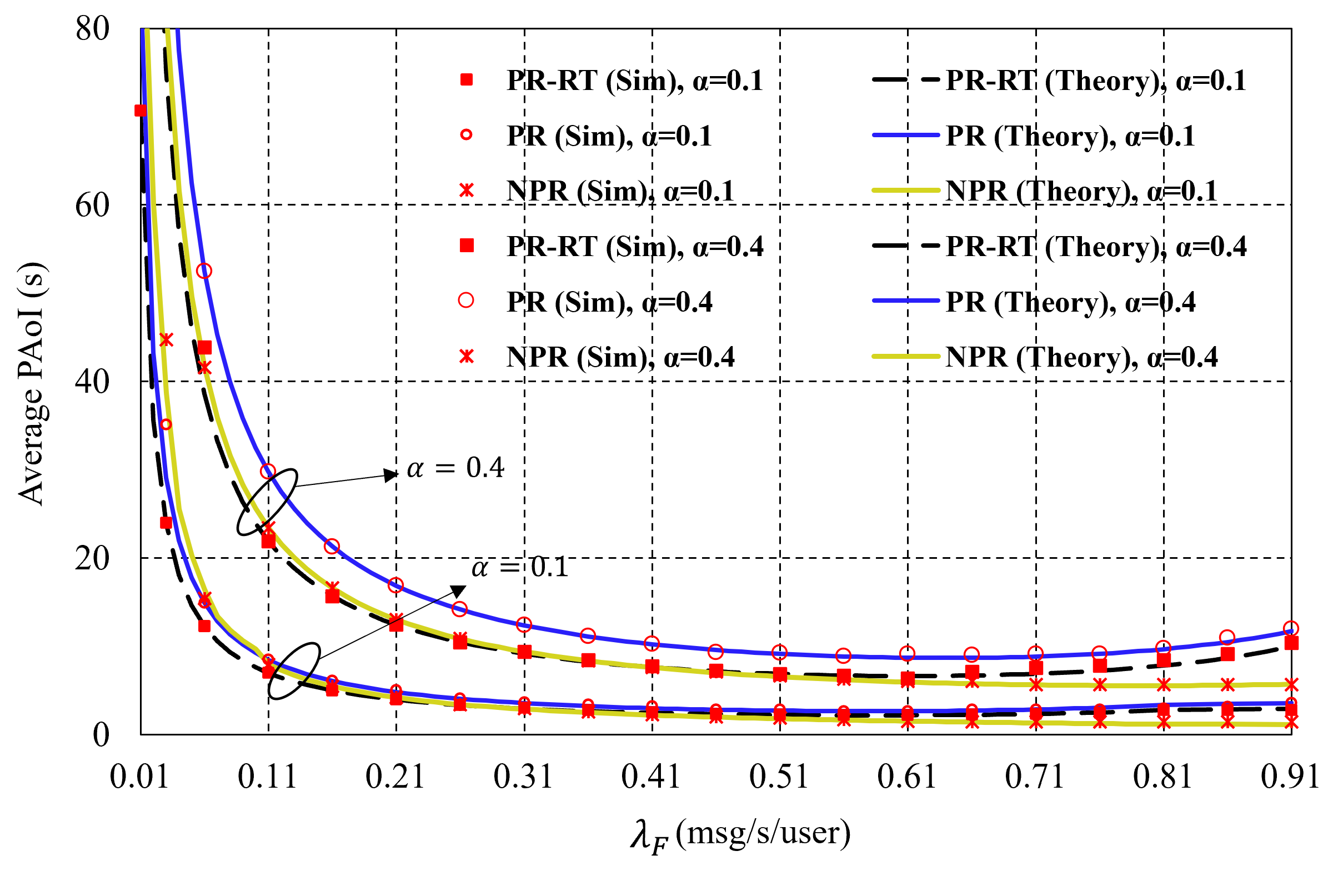}

		\caption{Analytical and simulation results of the average PAoI with varying $\lambda_F$ at $\mathrm{N_F}=10$ and $\alpha=0.1$ and 0.4 .  \label{theo-sim}}
	\end{figure}

Fig.~\ref{theo-sim} plots the average PAoI of obtained from the analytical models in Section~\ref{sec:model}, i.e., Eqs. \eqref{without_ret}, \eqref{with_re}, and \eqref{AoI-NP}, and the from the simulations under varying values of $\lambda_F$ and $\alpha$.
We can observe that the analytical results match well with the simulation ones, which validates the performed analysis in Section~\ref{sec:model}.
The figure also shows the impact of the retransmission mechanism in improving the average PAoI in PR-RT compared to PR.
This is mainly due to the fact that the PR-RT scheme utilizes idle transmission opportunities in the MCCH between packet generation instants to retransmit failed packets, which in turn reduces the average PAoI.
Moreover, it can be observed that although the average PAoI of the two schemes increases when $\alpha$ increases, the PR scheme experiences a sharper effect, especially under low values of $\lambda_F$.
This is because when $\alpha$ increases, it is more likely that the single transmission attempt would fail in the PR scheme, hence there would be a long waiting gap until the next packet is generated and received successfully.
Another important note from this figure is that the NPR scheme can introduce improved PAoI performance for relatively high values of $\lambda_F$ compared to the PR-RT one, which can be further explained as follows. For high values of $\lambda_F$, it is likely that a packet is preempted in the PR-RT scheme before it is received at the remote agent, which in turn increases the interdeparture time $\E[Y]$. In the NPR scheme, a new update is immediately transmitted when the last update have finished service, i.e., reducing $\E[Y]$ at high $\lambda_F$ compared to PR-RT approach.
In summary, the NPR scheme would be preferred for application scenarios with relatively high values of $\lambda_F$, otherwise, the PR-RT would be a better option. In the following results, we plot the simulation results of the average PAoI of all schemes.
    \begin{figure}[t!] 
		\centering
		\includegraphics[width= 0.9\linewidth]{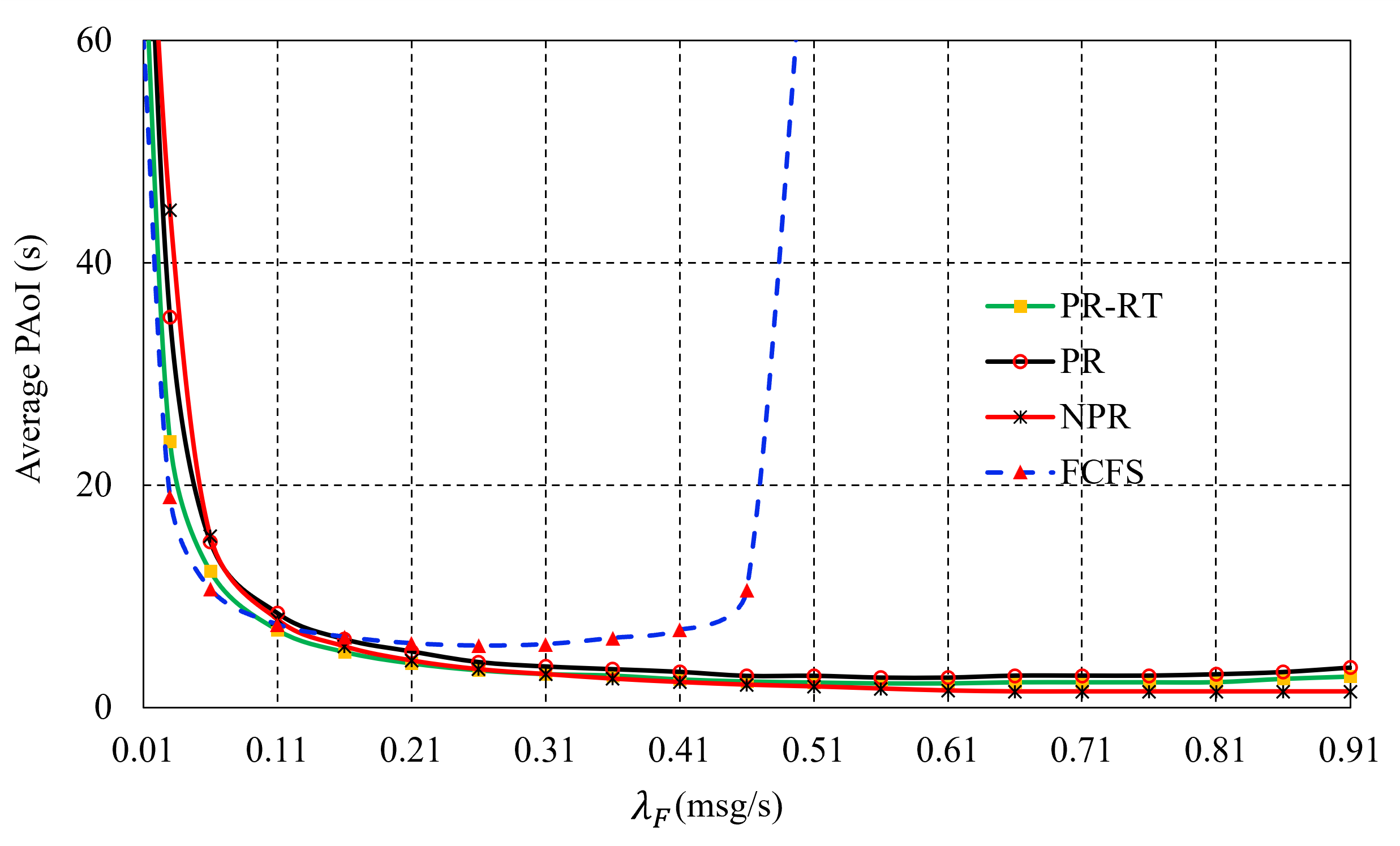}

		\caption{Comparison of the average PAoI under different queuing disciplines with $\mathrm{N_F}=10$ and $\alpha=0.1$.   \label{FCFS_comp}}
	\end{figure}

Fig.~\ref{FCFS_comp} compares the average PAoI of NPR, PR, PR-RT schemes and that of the standard \hl{packet-management} discipline in TETRA which is based on the FCFS approach with infinite buffer.
As $\lambda_F$ increases, the average PAoI first decreases in the FCFS scheme up to a certain value of $\lambda_F$, as the source tends to send updates more frequently.
However, beyond that critical value, the FCFS discipline incurs increase in the average PAoI due to the extended queuing length where the most recent packet spends long time in the output queue waiting while the
others in front are transmitted. 
This effect is, however eliminated in the queuing management schemes of NPR, PR and PR-RT as the queuing time no longer contributes to the average PAoI.
For very low values of $\lambda_F$, we can observe that the FCFS scheme outperforms NPR, PR and PR-RT schemes, which can be explained as follows.
These three schemes feature a queue size of just a single packet; when the interarrival times are large (i.e., low $\lambda_\text{F}$) there is a gap from the reception of the recent update until a new update is generated. In case of FCFS discipline, the transmitter is likely to be busy transmitting queued packets while newly generated ones arrive to the queue. 
For instance, with $\lambda_F=0.06$, the average PAoI of FCFS is $20\%$ less than that of PR-RT, while it is $90\%$ more than that of PR-RT when $\lambda_F$ increases to 0.3.
Therefore, for a optimal AoI performance, FCFS would be preferred for reporting rates up to 0.1~msg/s, and beyond that value a \hl{packet-management} scheme would be a better solution for improved AoI performance, especially for reporting rates beyond 0.4~msg/s.
    \begin{figure}[t!] 
		\centering
		\includegraphics[width= 1\linewidth]{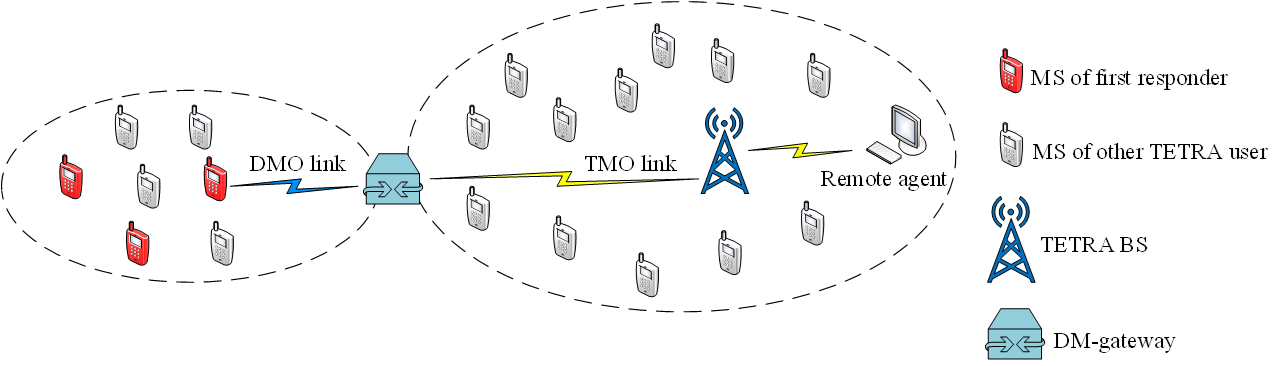}
		\caption{Network architecture of the simulated TETRA DMO model.  \label{netw-DMO}}
			\vspace*{-5mm}
	\end{figure}

\subsection{Performance of DMO and TMO service combined}

The DM-gateway communicates with the MSs of first responders over DMO links, then it retransmits the received information to the TETRA BS over TMO links as depicted in Fig.~\ref{netw-DMO}. We evaluate the performance of the DMO scenario under three different \hl{packet-management} settings as defined in Table.~\ref{settings}. In Fig.~\ref{DMO1}, we show the average PAoI for setting~(1) and setting~(2) with $\mathrm{N_F} \in \{ 5,10 \}$. In setting~(1), the first responders apply PR-RT while the DM-gateway applies FCFS with infinite buffer. In setting~(2), all MSs adopt the FCFS approach. Intuitively, the average PAoI increases for both settings compared to the TMO setting due to the additional transmission hop and the corresponding DM random channel access between the first responder and the DM-gateway. However, the PR-RT still manages to improve the average PAoI compared to FCFS for a long range of $\lambda_F$.
Since the DM-gateway adopts FCFS in setting~(1), the average PAoI first decreases as $\lambda_F$ increases, but then it tends to increase as $\lambda_F$ surpasses a certain value.
When a new  update is generated at the first responder adopting the PR-RT approach, it will preempt the current transmission at the first responder side. However, when it is received by the DM-gateway, it will still have to be queued at the FCFS-based buffer waiting for packets in front to be transmitted, which in turn increases the PAoI.
Also, when $\mathrm{N_F}$ increases, it is likely that users suffer from collisions, hence the average PAoI degrades for both settings, as depicted in Fig.~\ref{DMO1}.
For $\mathrm{N_F}=10$, the average PAoI can be kept below 60~s if $\lambda_F$ is roughly selected within the range 0.03-0.21~msg/s for setting~(2) while this range can be extended to be up to 0.81~msg/s when adopting setting~(1).   

Adopting NPR, PR or PR-RT approach at the DM-gateway could be a solution to alleviate the PAoI increase at higher values of $\lambda_F$ in setting (1). However, since the DM-gateway may serve as a relay for a large number of users within the DMO network, it could lead to a significant batch of updates to be either dropped or preempted as the queuing system for those three scheme can accommodate only a single update. In setting~(3), we let the DM-gateway adopts the \hl{packet-management} discipline proposed in~\cite{queu-mm1}. The \hl{packet-management} discipline presented in~\cite{queu-mm1} implies a queuing system with a capacity of two packets: a packet waiting in the queue, and another packet that is being served. The packet that is waiting for transmission is replaced upon arrival of a newly generated one. This way, the packet loss effect due to preemption at the gateway can be alleviated to a certain extent. Fig.~\ref{gateway-mm2} shows the average PAoI comparison between the three settings mentioned in~Table~\ref{settings}. As we can see in Fig.~\ref{gateway-mm2}, the 2-packet queuing approach in setting (3) can eliminate the effect of prolonged queuing length at the DM-gateway in setting~(1) and setting~(2) under high values of $\lambda_F$, which further improves the average PAoI.  
    \begin{figure}[t!] 
		\centering
		\includegraphics[width= 0.9\linewidth]{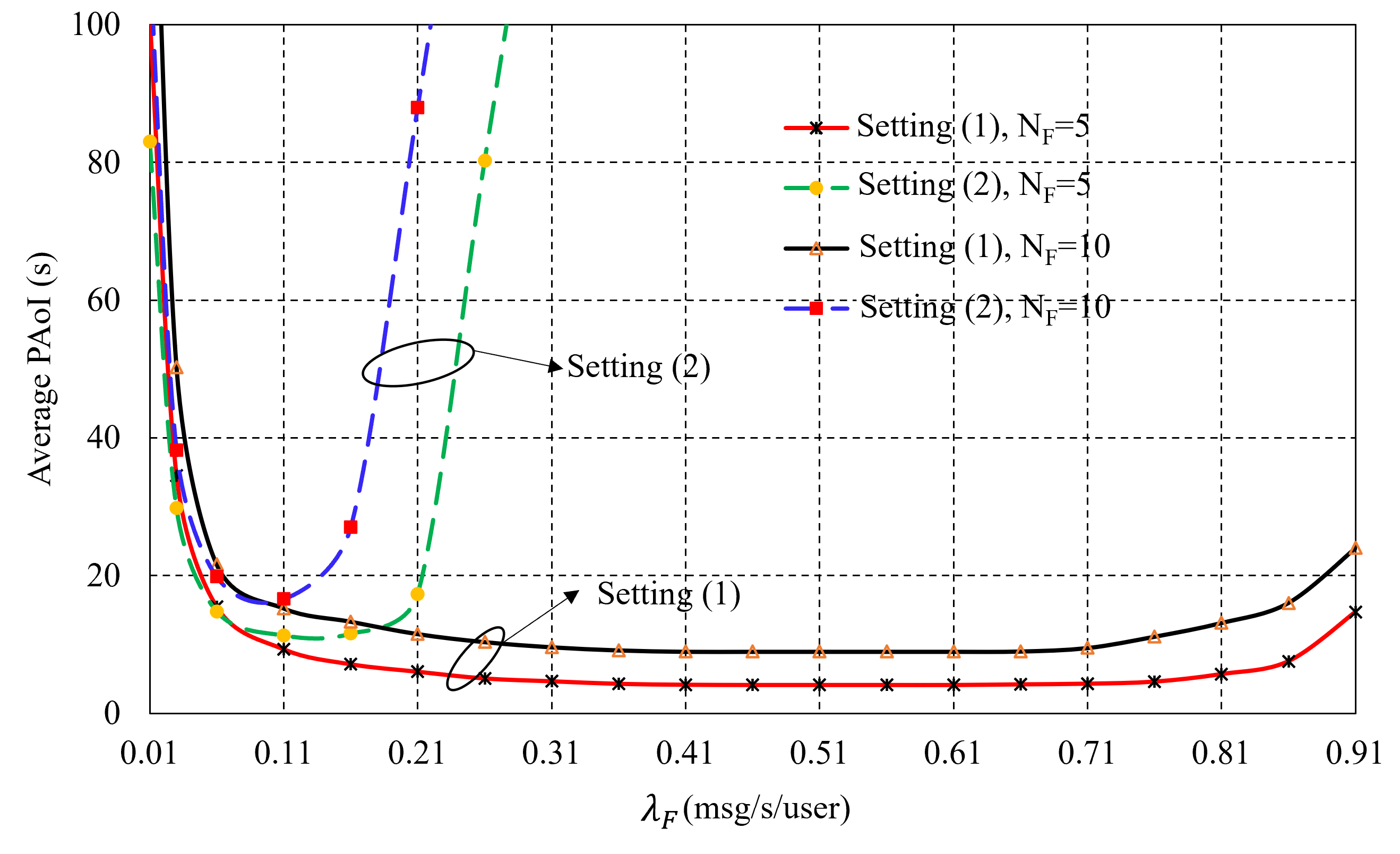}
		\caption{The average PAoI of setting~(1) and setting~(2) with $\alpha=0.1$ and $\mathrm{N_F} \in \{ 5,10 \}$.  \label{DMO1}}
	\end{figure}
    \begin{figure}[t!] 
		\centering
		\includegraphics[width= 0.9\linewidth]{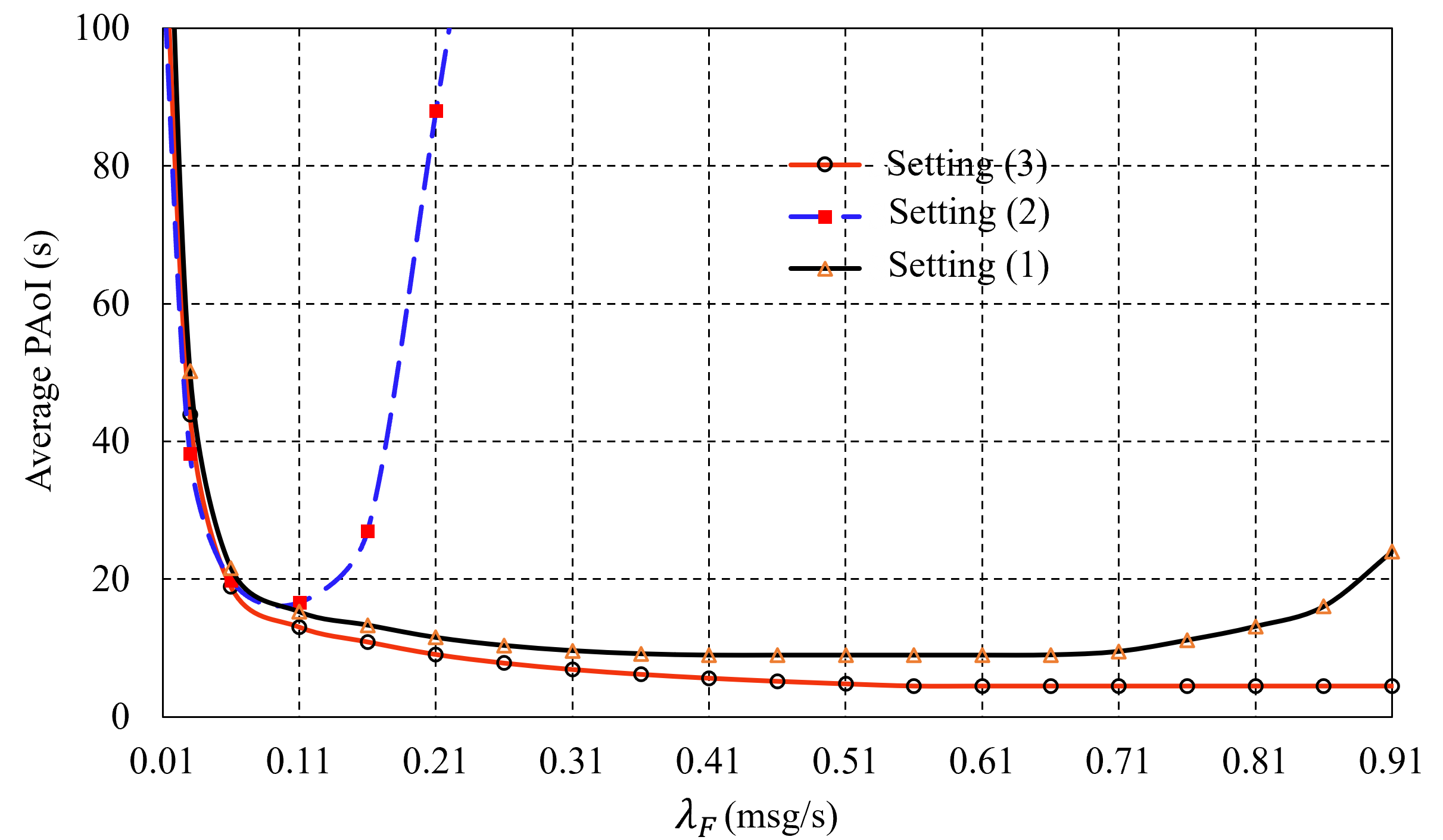}
		\caption{The Average PAoI under different \hl{packet-management} disciplines with $\alpha=0.1$ and $\mathrm{N_F}=10$.  \label{gateway-mm2}}
	\end{figure}
    \begin{figure}[t!] 
		\centering
		\includegraphics[width= 0.9\linewidth]{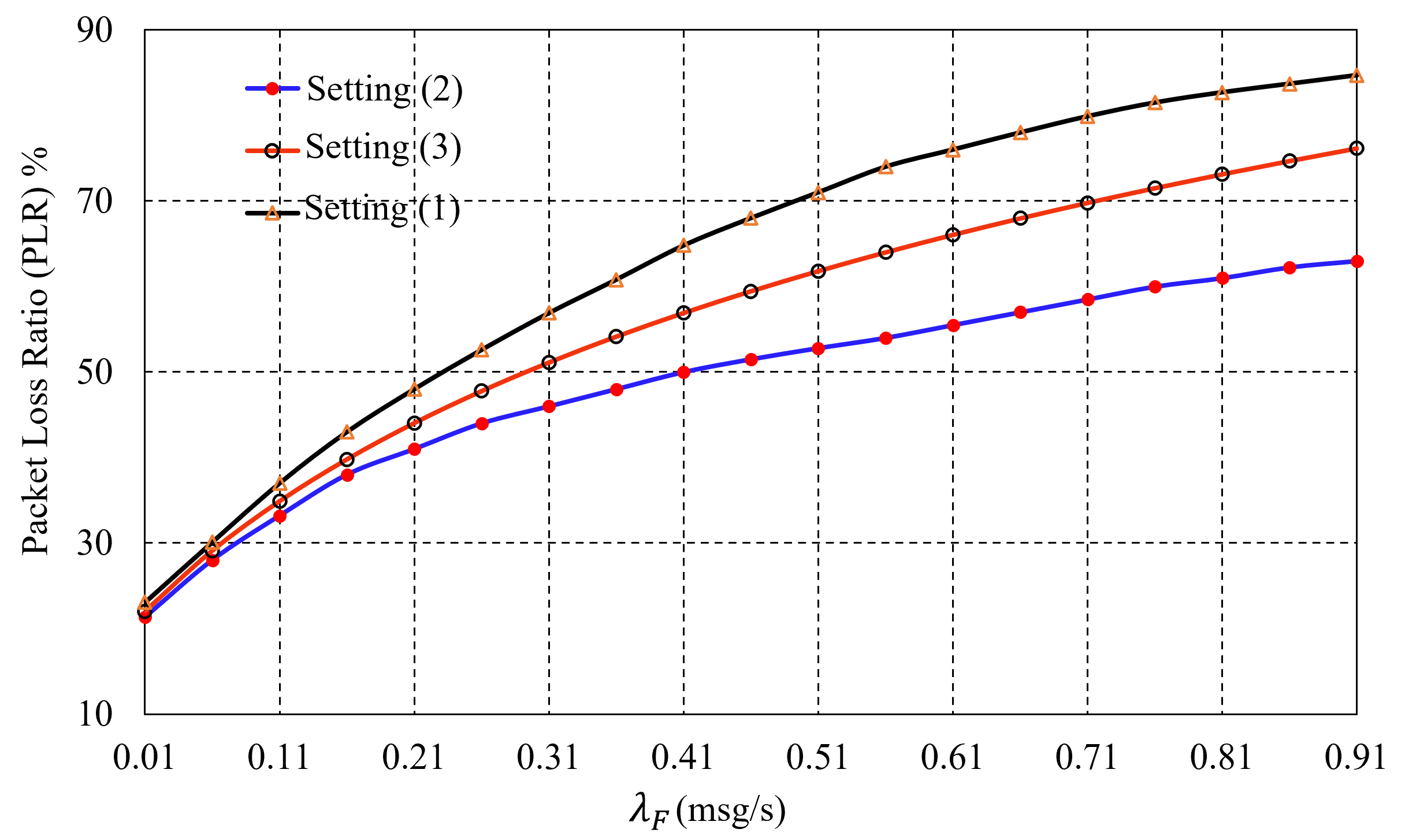}
		\caption{PLR comparison with varying $\lambda_F$ with $\mathrm{N_F}=10$.  \label{PDR}}
	
	\end{figure}

Finally, in Fig.~\ref{PDR}, we evaluate the packet loss performance of the three settings in terms of the Packet Loss Ratio (PLR).  \hl{Besides the fraction of packets lost due to the channel loss, the PLR parameter also accounts for the fraction of packets dropped due to preemption. Therefore, the results in Fig.~12 further demonstrate the trade-off between the AoI performance and the packet loss.}
The \hl{packet-management} scheme adopted in setting~(1) and setting~(3) leads to increased PLR compared to that of the FCFS in setting~(2), especially for higher values of $\lambda_F$ as illustrated by Fig.~\ref{PDR}, which can be explained as follows.
Since the first responders adopt PR-RT scheme in setting~(1), more packets (being served) are likely to be preempted the newly generated ones at high values of $\lambda_F$. In setting~(3), besides the batch of packets preempted at the first responders' queues, newly arriving packets at output buffer of the the DM-gateway are likely to be discarded due to the limited capacity of the two-packet queuing scheme~\cite{queu-mm1}.
On the other hand, the FCFS approach in setting~(2) accepts almost every generated packet in its infinite buffer, leading to lower PLR compared to that of setting~(1) and setting~(3).
We can also observe that the increased PLR in setting~(1) and  setting~(3) is insignificant for low values of $\lambda_F$ where the PLR is mainly affected by the radio-channel loss.
From Fig.~\ref{gateway-mm2} and  Fig.~\ref{PDR}, it may be stated that the increased PLR performance in  setting~(1) and setting~(3) would be a reasonable cost for improving the AoI performance, where the information timeliness could be more important than the link-level reliability.
For instance, at $\lambda_F=0.21$, the \hl{packet-management} approach in setting~(3) improves the average PAoI by $89\%$ compared to setting~(2) while the PLR increases by $8\%$ under the same value of $\mathrm{N_F}$.       

\section{Conclusion}
\label{sec:conclusions}

This work presented a proof-of-concept study of using TETRA as a connectivity enabler for remote monitoring of the physiological status of first responders.
Specifically, we have investigated the communication performance of TETRA SDS in terms of the PAoI metric as a measure of the freshness of the application messages exchanged in the system.
The results showed that adopting \hl{packet-management} schemes has a favorable effect on the PAoI performance with a modest increase in the packet loss ratio.
The results also provide insights on the expected performance of TETRA SDS in the considered remote-health monitoring scenario, which further help in optimizing the key system parameters for the target performance requirements. \hl{As a future work, the investigations in this paper could be extended to include a comparison of the AoI performance of the health-monitoring system  considering different communication technologies, e.g., LTE or 5G, where the corresponding standards are open  to adopt more advanced link scheduling and packet-management policies.}
\section*{Acknowledgement}
This paper has received funding from the European Union’s Horizon 2020 research and innovation programme under grant agreement No. 883315.

	\bibliographystyle{IEEEtran}
\bibliography{mybib}

\end{document}